\documentclass[11pt,a4paper]{article}
\usepackage[margin=28mm]{geometry}
\usepackage{amsmath,amssymb,amsthm}
\usepackage{microtype}
\usepackage{tikz}
\newtheorem{theorem}{Theorem}
\newtheorem{lemma}{Lemma}
\newcommand{\Top}{\operatorname{Top}}

\title{Breaking the Square-Root Space Barrier for Longest Increasing Subsequence}
\author{
  Naohito Matsumoto \quad Masashi Kiyomi\\[1ex]
  \normalsize Seikei University
}
\date{}
\begin{document}
\maketitle
\begin{abstract}
For every fixed constant $\varepsilon>0$, we give deterministic algorithms for computing the length of a longest increasing subsequence (LIS) and for outputting an LIS in input order, both in ${\rm O}(n^{2+\varepsilon})$ time and ${\rm O}(\sqrt n/2^{\delta\sqrt{\log n}})={\rm o}(\sqrt n)$ bits of working space for some constant $\delta>0$. The input is read-only and the output is write-only. This answers an open question of Kiyomi et al. Our construction combines small-pile skipping with recursive blockwise computation of pile tops. Balanced reconstruction uses streaming intersections of forward and backward piles.
\end{abstract}
\section{Introduction}
The longest increasing subsequence (LIS) problem is a classical problem in algorithm design. Given a sequence $A=(a_1,\ldots,a_n)$ of keys from a totally ordered domain, the task is to find a longest subsequence $a_{i_1},\ldots,a_{i_\ell}$ such that $i_1<\cdots<i_\ell$ and $a_{i_1}<\cdots<a_{i_\ell}$. The length of an LIS can be computed in ${\rm O}(n\log n)$ time by maintaining, for each attainable subsequence length, the smallest possible last key. This approach, often described in terms of patience sorting~\cite{aldous,bespamyatnikh,fredman}, uses ${\rm O}(n)$ words of working memory in the worst case. In patience sorting, keys are processed from left to right. Each key is placed on the leftmost pile whose top key is larger than it (assuming distinct keys), or on a new pile if no such pile exists. After processing $A$, the top of the $k$th pile is the smallest possible last key of an increasing subsequence of length $k$ in $A$. We call this value the \emph{pile top} $p_k(A)$. The number of nonempty piles equals the LIS length. In this paper we study how much this space requirement can be reduced while retaining polynomial running time.
The precise model is given in Section~2.

The decision version of LIS, which asks whether an increasing subsequence of length at least $k$ exists, belongs to $\mathrm{NL}$: a nondeterministic machine guesses successive positions, retaining only the previous position and a counter. For keys represented by ${\rm O}(\log n)$ bits, this requires ${\rm O}(\log n)$ bits of space. By Savitch's theorem~\cite{savitch}, the decision problem, and hence the LIS length by binary search on $k$, can be solved deterministically in ${\rm O}(\log^2 n)$ bits of space and $n^{{\rm O}(\log n)}$ time. This does not, however, guarantee polynomial running time. Our focus is on simultaneously achieving polynomial time and small deterministic workspace.

Time-space trade-offs have been extensively studied for fundamental algorithmic problems.
For example, Munro and Paterson~\cite{MP80} studied selection and sorting with limited storage.
Munro and Raman~\cite{MR96} developed selection algorithms using only a constant number of extra storage cells and ${\rm O}(n^{1+\varepsilon})$
time for any fixed $\varepsilon>0$. Barnes et al.~\cite{BBRS98} gave a deterministic polynomial-time algorithm for directed $s$-$t$ connectivity using $n/2^{\Theta(\sqrt{\log n})}$ space. In the context of dynamic programming, Hirschberg's classical algorithm~\cite{H75} computes a longest common subsequence in quadratic time using linear working space in words. More recently, Kiyomi, Horiyama, and Otachi~\cite{lcs} obtained a polynomial-time sublinear-space algorithm for LCS, whose recursive blockwise approach also inspires our algorithm.

Kiyomi et al.~\cite{lisjournal} gave a deterministic algorithm that computes the LIS length in ${\rm O}(n^2\log n/s)$ time using ${\rm O}(s\log n)$ bits of working space, for $\lceil\sqrt n\rceil\le s\le n$. They also gave an algorithm for outputting an LIS with an additional logarithmic factor in time. Their length algorithm exploits the partition of the input into piles according to the length of a longest increasing subsequence ending at each position. By storing a small pile, the algorithm can resume its computation at a higher rank without retaining all preceding piles. The counting argument ensuring that another small pile can be reached within the available workspace, however, requires $s\ge\sqrt n$. In particular, this trade-off does not provide an algorithm using ${\rm o}(\sqrt n)$ bits.
In Section 7 of~\cite{lisjournal}, the authors explicitly
asked whether LIS can be solved in polynomial time using
${\rm o}(\sqrt n)$ space.
As also observed in~\cite[Section 7]{lisjournal},
a negative answer to this question would separate
$\mathrm{SC}$ from
$\mathrm{P}\cap\mathrm{PolyL}$.

\paragraph{Our results.}
We answer this question affirmatively. For every fixed constant $c>0$, we give deterministic algorithms for computing the LIS length and for outputting an LIS in input order, both running in $n^{2+4c^2+{\rm o}(1)}$ time and using
\[
 {\rm O}\left(\frac{\sqrt n(\log n)^{3/2}}{2^{c\sqrt{\log n}}}\right)
\]
bits of working space. In particular, for every fixed $\varepsilon>0$, there exists a constant $\delta>0$ such that both tasks can be performed in ${\rm O}(n^{2+\varepsilon})$ time and ${\rm O}(\sqrt n/2^{\delta\sqrt{\log n}})$ bits of working space. The space improvement over $\sqrt n$ is asymptotically larger than any fixed polylogarithmic factor, although it does not reduce the exponent of $n$ by a positive constant.

\paragraph{Technical overview.}
Our construction is inspired by the polynomial-time sublinear-space algorithm for longest common subsequence (LCS) of Kiyomi, Horiyama, and Otachi~\cite{lcs}. That algorithm processes blocks of boundary values in a recursive decomposition of a directed grid. For LIS, a corresponding difficulty is to summarize the information about increasing subsequences in the first half of the input without storing the entire array of minimum last keys (tops). Merely applying the earlier small-pile scanning algorithm with a smaller workspace does not resolve this difficulty: the number of consecutive large piles may exceed the number of tops that can be maintained simultaneously.

The main ingredient of our algorithm is a block-combination lemma for tops. After splitting the input into two consecutive parts, we partition the relevant ranks of the first part into blocks. Each block of finite tops is itself an increasing sequence. We use it as an artificial prefix of a suitably filtered second part and adjust the requested ranks. The tops of the original sequence can then be recovered by taking componentwise minima over the blocks. Crucially, every resulting subproblem is again a top query on an input interval with a short artificial prefix and a single key threshold. Thus we can split the original interval recursively while keeping the workspace of each recursion frame small. When the interval is sufficiently short, a truncated variant of the earlier LIS algorithm can be used directly.

To keep the branching factor of this recursion small, we retain the small-pile traversal of~\cite{lisjournal}. Rather than maintaining all tops in a potentially large rank window, we enumerate the members of individual piles by binary searches on prefix tops. The disjointness of the piles bounds the total number of such searches. Balancing the workspace per recursion frame against the branching factor and recursion depth yields the stated time--space bounds.

For outputting an actual LIS, we first observe that successive endpoint computations give a simple reconstruction algorithm at the cost of an additional factor of $n$ in time. We then remove this factor by adapting the forward/backward pile-intersection framework of~\cite{lisjournal}. Our top queries allow arbitrary piles to be enumerated in either position order without storing them in their entirety. Two such enumerations can be merged to find a middle element of an LIS, permitting balanced recursive reconstruction within the same asymptotic resource bounds as length computation.

\begin{theorem}[Main theorem]\label{thm:main}
For every fixed $\varepsilon>0$, there exists a constant $\delta>0$ such that the length of an LIS of a read-only sequence of length $n$ can be computed, and an LIS can be output in input order to a write-only output, deterministically in ${\rm O}(n^{2+\varepsilon})$ time and
\[
{\rm O}\left(\frac{\sqrt n}{2^{\delta\sqrt{\log n}}}\right)={\rm o}(\sqrt n)
\]
bits of working space.
\end{theorem}
The length bound is established in Theorem~\ref{thm:length}, and the output bound in Theorem~\ref{thm:fastoutput}. Choosing $4c^2<\varepsilon$ and $0<\delta<c$ in those bounds proves the main theorem.

\section{Model and notation}
All logarithms are base two. We work in a logarithmic-word model, where each word
contains $\Theta(\log n)$ bits and arithmetic on word-sized quantities takes constant
time. The input is a read-only array of $n$ values with random access, and, when an
actual subsequence is required, the output is written to a write-only tape. Space bounds
count only the working space, in bits. A key is represented by the position (index) of
its input value, which takes ${\rm O}(\log n)$ bits. We assume that the input values are
integers of ${\rm O}(\log n)$ bits, so comparing two keys, which reads the two corresponding
input values, takes constant time. (For $w$-bit values, every running time below is
multiplied by at most $\lceil w/\log n\rceil$, while the space bounds are unchanged,
since values are never stored.)
For strict LIS with repeated keys, order keys lexicographically by (value, negative
original position), thereby making them distinct without changing increasing
subsequences. Artificial keys below retain their original identities.

For a sequence $A$, we define the pile top $p_k(A)$ formally as the minimum last key of an increasing subsequence of length $k$, or $+\infty$ if none exists. Set $p_0(A)=-\infty$. The finite positive-index tops form a strictly increasing sequence. Equivalently, $p_k$ for $k>0$ is the minimum endpoint of a chain of length at least $k$, since a suffix of such a chain retains its endpoint.

For a sequence under consideration, the \emph{absolute rank} of a position is the length of a longest increasing subsequence ending there, and its pile $P_i$ consists of all positions of absolute rank $i$. An \emph{anchor pile} is a pile retained in memory to resume computation at higher ranks. Relative ranks are measured in the sequence eligible with respect to this anchor; Lemma~\ref{lem:skip} shows that relative rank $j$ corresponds to absolute rank $i+j$ when the anchor has rank $i$. A pile top in either sequence is defined by the same function $p_k$; the sequence argument determines which rank is meant.

\section{The earlier limited-workspace LIS algorithm}\label{sec:stacs}
\begin{theorem}[Kiyomi et al.~\cite{lisjournal}]\label{thm:stacs}
For $\lceil\sqrt M\rceil\le u\le M$, LIS length is computable in ${\rm O}(M^2\log(2M)/u)$ time using ${\rm O}(u\log(2M))$ bits in the usual logarithmic-word model.
\end{theorem}
We review the length algorithm of Kiyomi et al.~\cite{lisjournal}, translating its notation to ours. We also give a self-contained justification of the filtering property for use in the new algorithm. Let $B$ be a sequence of $M$ distinct keys and let $P_i(B)$ contain the positions whose longest increasing subsequence ending there has length $i$. Each arrival updates exactly one pile top: a new key $x$ improves $p_i$ precisely when $p_{i-1}<x<p_i$. Thus each pile is decreasing in input order.

\paragraph{Scanning from a known pile.}
Suppose $P_i$ is stored in input order. Ignore its own elements and all keys that do not exceed its most recent preceding member; before the first member, ignore every key. For $i=0$, retain everything. Lemma~\ref{lem:skip} below proves that the retained sequence has ranks shifted by $i$. Maintain only the first $h$ relative tops, initially infinite, and one counter per top. On reading a retained key, locate the first top exceeding it by binary search. If there is no such top, discard the key; otherwise replace that top and increment its counter. Higher ranks cannot influence lower tops. The counters give $|P_{i+1}|,\ldots,|P_{i+h}|$. If some counter is zero, the first such rank determines the LIS length. A second scan can collect the positions assigned to any selected rank. This is the mechanism used in~\cite{lisjournal}.

\paragraph{Choosing the next pile.}
For a parameter $u$ satisfying $\lceil\sqrt M\rceil\le u\le M$, take $h=2u$ in each scan. Starting at $i=0$, terminate if the last examined pile is empty. Otherwise, among the $u$ piles numbered $i+u+1,\ldots,i+2u$, at least one has size at most $M/u\le u$, since all piles are disjoint. Collect that pile in the second scan, discard the previous one, and repeat from its rank. Each iteration advances by at least $u$ or terminates. With constant-time access to keys, each scan costs ${\rm O}(M\log(2u))$ time, and the stored piles, tops, and counters occupy ${\rm O}(u)$ words.

In our representation keys are original input indices, so a word requires ${\rm O}(\log n)$ bits rather than necessarily ${\rm O}(\log M)$ bits. The same procedure therefore uses ${\rm O}(u\log n)$ bits. The cases $M=0,1$ are immediate. The earlier paper also establishes a witness-output algorithm with an additional logarithmic time factor (Theorem 5.1), whose divide-and-conquer framework we use in Section~\ref{sec:witness}, and a sequential-access time--space lower bound (Theorem 6.2).

\paragraph{Where the square-root restriction arises.}
The traversal above can store $u$ complete pile entries and ${\rm O}(u)$ simultaneous tops. To ensure a small pile in a window of $u$ ranks, the counting argument requires $M/u\le u$. If $u<\sqrt M$, a run of more than $u$ large piles may occur, so that direct truncated patience sorting cannot guarantee the next jump with this workspace. This is a limitation of the construction, not a space lower bound for LIS. Our algorithm retains the small-pile traversal but replaces the simultaneous maintenance of a potentially wide rank window by recursive batched queries. Once the input interval has length at most $s^2$, the earlier algorithm again fits in the available workspace.

\section{Combining blocks of pile tops}
Our goal is to compute a requested range of pile tops without maintaining all pile tops simultaneously. Split the input sequence $A$ by position into a prefix $X$ and a suffix $Y$. If all pile tops of $X$ were available, ordinary patience sorting could continue on $Y$ from that state. Instead, we partition the \emph{rank indices} of the pile tops of $X$ into small blocks. Each block provides a short increasing sequence, used as an artificial prefix of a subsequence of $Y$. The following lemma shows how to combine the results of these smaller computations. Notice that the split into $X$ and $Y$ is by input position, whereas the blocks are by subsequence length.

\begin{lemma}\label{lem:block}
Let $X$ and $Y$ be sequences of distinct keys with disjoint key sets, and let $A=XY$ denote their concatenation. Let $r,s\ge 1$ be integers. Partition the rank indices $0,\ldots,r$ into consecutive blocks of size at most $s$. For a block $[a,b]$ with $a>0$ and $p_a(X)<+\infty$, let $C$ be the sequence $(p_a(X),\ldots,p_b(X))$ with infinite entries removed, and let $Y'$ consist of the keys in $Y$ greater than $p_a(X)$. For every $k\in[1,r]$ with $k\ge a$, the block contributes the candidate $p_{k-a+1}(CY')$. Blocks with $p_a(X)=+\infty$ or $a>k$ contribute $+\infty$. For the block starting at zero, let $C$ consist of its finite positive-index tops; its contribution is $p_k(CY)$. Then $p_k(A)$ is the minimum of all block contributions.
\end{lemma}
\begin{proof}
Consider a chain of length exactly $k$ in $XY$ attaining $p_k(A)$, and let $i$ be the number of its keys in $X$. If $i>0$, replace its $X$ part by a chain of length $i$ ending at $p_i(X)$. Its endpoint does not increase, and the $Y$ part remains feasible. In the block containing $i$, the artificial prefix through $p_i(X)$ represents this chain with length reduced by $a-1$ when $a>0$, and unchanged when $a=0$. All used $Y$ keys survive the filter. If $i=0$, the block starting at zero represents the chain directly. Thus some candidate is at most $p_k(A)$.

Conversely, consider a chain attaining a finite candidate. If its last artificial key is $p_i(X)$, insert all earlier keys of that artificial prefix and replace the resulting prefix by an actual length-$i$ chain in $X$. For $a>0$ this increases length by at least $a-1$ relative to the original artificial chain; for $a=0$ it does not decrease length. The endpoint is unchanged. If the chain uses no artificial key and $a>0$, prepend a length-$a$ chain ending at $p_a(X)$, possible by the filter. For $a=0$, no such operation is needed. In every case the resulting chain has length at least $k$ and the same endpoint. Hence every candidate is at least $p_k(A)$.
\end{proof}

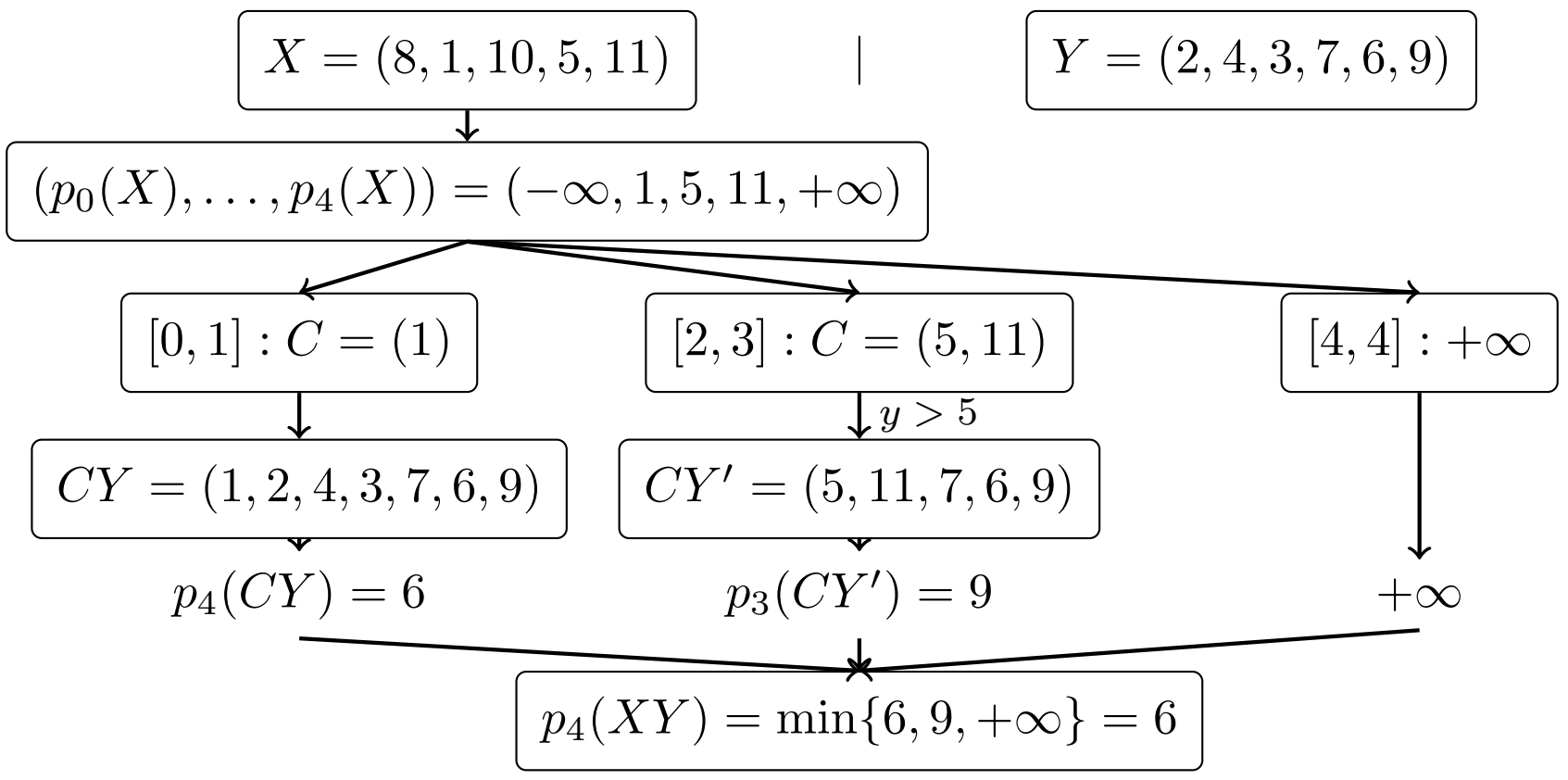
\begin{figure}[htbp]
\centering
\begin{tikzpicture}[x=1cm,y=0.72cm, every node/.style={font=\small}, box/.style={draw,rounded corners=2pt,minimum height=0.55cm,inner sep=5pt}, arr/.style={->,thick}]
\node[box] (x) at (-2.8,0) {$X=(8,1,10,5,11)$};
\node[box] (y) at (2.8,0) {$Y=(2,4,3,7,6,9)$};
\node at (0,0) {$\mid$};
\node[box] (tops) at (-2.8,-1.3) {$(p_0(X),\ldots,p_4(X))=(-\infty,1,5,11,+\infty)$};
\draw[arr] (x) -- (tops);
\node[box] (b0) at (-4,-2.8) {$[0,1]: C=(1)$};
\node[box] (b1) at (0,-2.8) {$[2,3]: C=(5,11)$};
\node[box] (b2) at (4,-2.8) {$[4,4]: +\infty$};
\foreach \b in {b0,b1,b2} \draw[arr] (tops.south) -- (\b.north);
\node[box] (q0) at (-4,-4.25) {$CY=(1,2,4,3,7,6,9)$};
\node[box] (q1) at (0,-4.25) {$CY'=(5,11,7,6,9)$};
\draw[arr] (b0) -- (q0);
\draw[arr] (b1) -- node[right,font=\scriptsize] {$y>5$} (q1);

\node (v0) at (-4,-5.3) {$p_4(CY)=6$};
\node (v1) at (0,-5.3) {$p_3(CY')=9$};
\node (v2) at (4,-5.3) {$+\infty$};
\draw[arr] (q0)--(v0);\draw[arr] (q1)--(v1);\draw[arr] (b2)--(v2);
\node[box] (out) at (0,-6.55) {$p_4(XY)=\min\{6,9,+\infty\}=6$};
\foreach \v in {v0,v1,v2} \draw[arr] (\v.south)--(out.north);
\end{tikzpicture}
\caption{Computing a pile top by splitting the input by position and the pile-top indices by rank. The blocks are processed one at a time; the diagram shows all three for illustration.}
\label{fig:blocks}
\end{figure}

\paragraph{Example.}
Let $X=(8,1,10,5,11)$ and $Y=(2,4,3,7,6,9)$, and suppose we seek $p_4(XY)$ with block size $s=2$. The finite pile tops of $X$ are $(1,5,11)$, and the blocks of ranks are $[0,1]$, $[2,3]$, and $[4,4]$. Figure~\ref{fig:blocks} illustrates the computation. The first block gives the artificial sequence $(1,2,4,3,7,6,9)$ and candidate $p_4=6$. For the second block, $C=(5,11)$ and the threshold is $5$, so $Y'=(7,6,9)$. Its candidate is $p_{4-2+1}(5,11,7,6,9)=9$. The third block has an infinite first top and contributes $+\infty$. Taking the minimum gives $p_4(XY)=6$ attained, for example, by $(1,2,4,6)$.

\section{Recursive batched top queries}
Fix a workspace parameter $s\ge 1$ and a maximum requested rank $d$. The arguments of $\Top(C,I,v,l,r)$ are as follows:
\begin{itemize}
\item $C$: an artificial increasing prefix of at most $s$ keys;
\item $I$: an interval of positions in the original input;
\item $v$: a lower key threshold, applied only to original interval keys;
\item $l,r$: the first and last requested ranks, with $0\le l\le r\le d$ and $r-l+1\le s$.
\end{itemize}
Let $B$ be the concatenation of $C$ and the eligible keys of $I$ exceeding $v$, in position order. The query returns $(p_l(B),\ldots,p_r(B))$. Eligibility is fixed by the outer algorithm (Section~\ref{sec:outer}); rank zero is returned directly as $-\infty$.

\paragraph{Base case.} If $|I|\le s^2$, use a truncated version of the traversal in Section~\ref{sec:stacs}. The logical input $B$ consists of at most $s^2+s$ keys. Starting at rank zero, compute the next $3s$ relative tops and their pile sizes by a scan. Save all requested tops that fall in this window in the output array. If the window reaches the largest requested rank, or contains an empty pile, finish (unattained ranks are infinite). Otherwise select a pile of size at most $s$ among the last $2s$ ranks; such a pile exists since $2s(s+1)>s^2+s$. Collect its positions in a second scan and continue from that pile, advancing at least $s$ ranks. 

The shift property of Lemma~\ref{lem:skip} applies to the logical sequence, including its artificial prefix. The final tops in each relative window are exactly the corresponding absolute tops. Each scan reads the artificial prefix and the raw interval directly, testing the fixed outer eligibility by binary search in the original anchor pile and applying the threshold only to original interval keys. A second local eligibility test uses positions in the logical sequence (or sequential scan counters), not original identities. 

Each scan costs ${\rm O}((|I|+s)\log(2n))$ time and uses ${\rm O}(s\log n)$ bits. There are ${\rm O}(1+r/s)$ stages, so the leaf costs ${\rm O}((s^2+sd)\log(2n))={\rm O}(n\log(2n))$ time.

\paragraph{Recursive case.} Otherwise split $I$ in half. Write $X$ and $Y$ for the corresponding eligible, threshold-filtered sequences. The left input is $CX$ and the right input is $Y$. Initialize the output to $+\infty$ (except rank zero). Process the blocks of $0,\ldots,r$ one at a time:
\begin{enumerate}
\item Recursively compute the tops in $[a,b]$ for $CX$ using the left interval and the inherited prefix and threshold.
\item If $a>0$ and the first top is infinite, skip this block. Otherwise form the artificial prefix $C'$ from its finite positive-index tops.
\item If $a=0$, query the right interval with prefix $C'$, unchanged threshold, and the requested positive ranks. If $a>0$ and $a\le r$, query it with prefix $C'$, threshold $\max(v,p_a(CX))$, and ranks $\max(l,a)-a+1,\ldots,r-a+1$.
\item Map the returned ranks back to their original indices and take componentwise minima with the accumulated output.
\end{enumerate}
Figure~\ref{fig:rec} summarizes the two recursive calls for one block. The following pseudocode makes their order explicit; $R$ is the accumulated output array indexed by rank.
\begin{quote}\small
\textbf{Procedure} $\Top(C,I,v,l,r)$\\
\hspace*{1em} If $r=0$, return $(-\infty)$.\\
\hspace*{1em} If $|I|\le s^2$, return the base-case scan results.\\
\hspace*{1em} Split $I$ into $I_L,I_R$; initialize $R$ as above.\\
\hspace*{1em} \textbf{for each} block $[a,b]$ of $[0,r]$:\\
\hspace*{2em} $T\gets\Top(C,I_L,v,a,b)$\\
\hspace*{2em} If $a>0$ and $T[a]=+\infty$, continue.\\
\hspace*{2em} $C'\gets$ finite positive-rank entries of $T$ in rank order.\\
\hspace*{2em} \textbf{if} $a=0$:\\
\hspace*{3em} $U\gets\Top(C',I_R,v,\max(1,l),r)$\\
\hspace*{3em} $R[k]\gets\min(R[k],U[k])$ for $\max(1,l)\le k\le r$.\\
\hspace*{2em} \textbf{else}:\\
\hspace*{3em} $U\gets\Top(C',I_R,\max(v,T[a]),\max(l,a)-a+1,r-a+1)$\\
\hspace*{3em} $R[k]\gets\min(R[k],U[k-a+1])$ for $\max(l,a)\le k\le r$.\\
\hspace*{1em} Return $R$.
\end{quote}
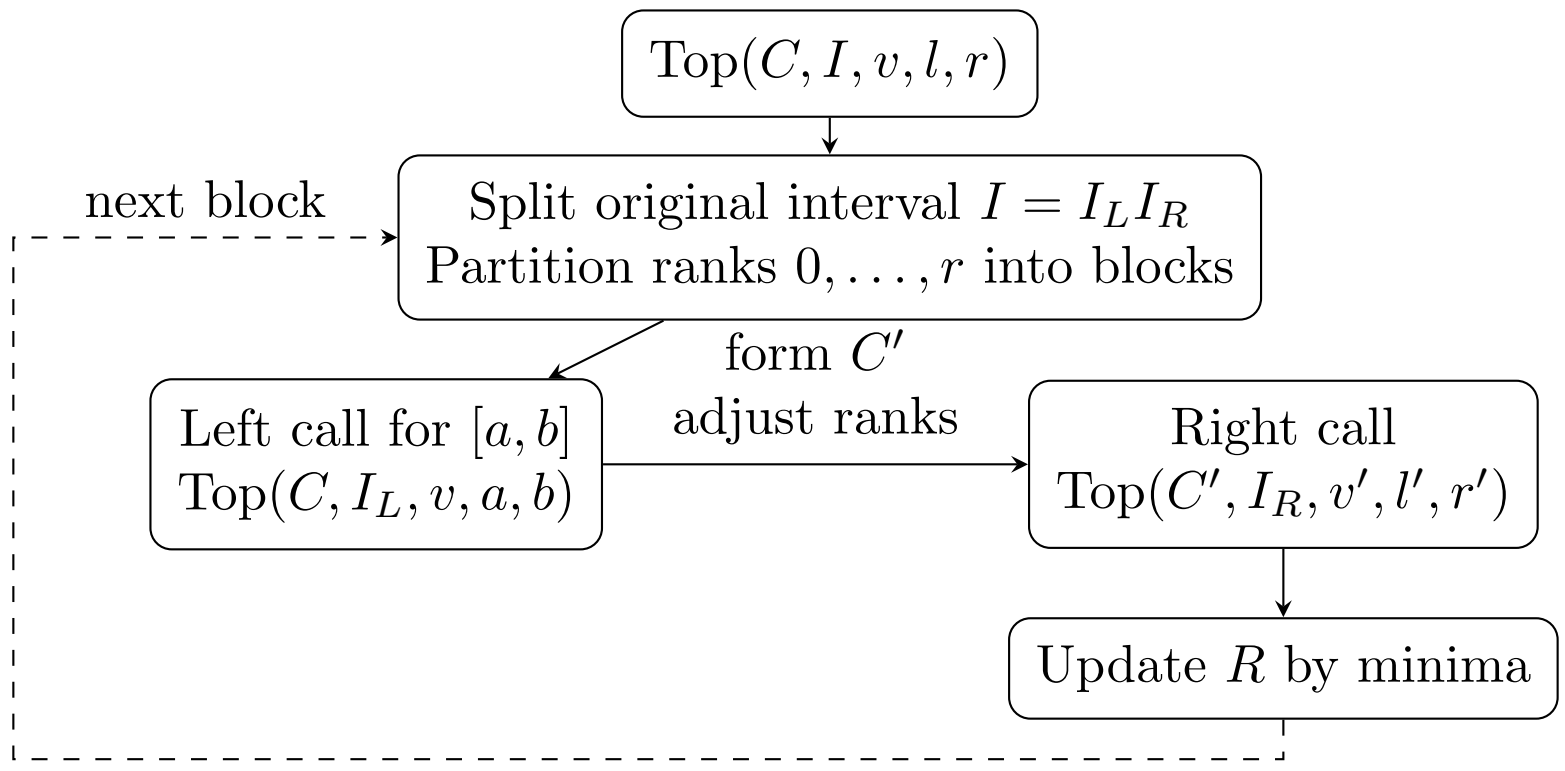
\begin{figure}[htbp]
\centering
\begin{tikzpicture}[every node/.style={font=\small}, box/.style={draw,rounded corners,align=center,inner sep=5pt},>=stealth]
\node[box] (root) at (0,0) {$\Top(C,I,v,l,r)$};
\node[box] (split) at (0,-1.15) {Split original interval $I=I_L I_R$\\ Partition ranks $0,\ldots,r$ into blocks};
\node[box] (left) at (-3,-2.65) {Left call for $[a,b]$\\ $\Top(C,I_L,v,a,b)$};
\node[box] (right) at (3,-2.65) {Right call\\ $\Top(C',I_R,v',l',r')$};
\node[box] (out) at (3,-4) {Update $R$ by minima};
\draw[->] (root)--(split); \draw[->] (split)--(left);
\draw[->] (left)--node[above,align=center]{form $C'$\\ adjust ranks} (right);
\draw[->] (right)--(out);
\draw[->,dashed] (out.south) -- (3,-4.6) -- (-5.4,-4.6) -- (-5.4,-1.15) -- node[above]{next block} (split.west);
\end{tikzpicture}
\caption{Recursive structure of a nonleaf query. The left call supplies the artificial prefix for the right call; $v',l',r'$ are determined by the block as in the pseudocode. Calls and blocks are processed sequentially. Infinite blocks may be skipped.}
\label{fig:rec}
\end{figure}
If there are no positive ranks to query, return immediately. Lemma~\ref{lem:block} and induction on interval size prove correctness. The largest requested rank never increases. Each new threshold is a single key, so no nested filtering representation is needed. The artificial prefix is replaced, rather than appended to an ancestral prefix, on a right call. Consequently each stack frame holds only ${\rm O}(s)$ keys and indices.

\section{Skipping small piles}\label{sec:outer}
Let $L(q)$ be the length of a longest increasing subsequence ending at input position $q$, and let $P_i=\{q:L(q)=i\}$. These sets partition the input. Within a pile, keys decrease in input order.
\begin{lemma}\label{lem:skip}
Given $P_i$ for $i>0$, retain precisely the positions $q$ for which some $p\in P_i$ precedes $q$ and has a smaller key. In the resulting sequence $A'$, the rank of every retained position is $L(q)-i$. Exactly the positions of rank greater than $i$ are retained.
\end{lemma}
\begin{proof}
Any increasing chain in $A'$ can be prefixed by a length-$i$ chain ending at a suitable member of $P_i$ before its first key. Thus its length ending at $q$ is at most $L(q)-i$. Conversely, take a longest chain ending at a position with $L(q)>i$. Its $i$th member has rank exactly $i$: a longer chain to that member would extend to a longer chain to $q$. Every later member is retained, giving a chain of length $L(q)-i$ in $A'$. These observations also prove the equivalence of eligibility and $L(q)>i$.
\end{proof}

Initially set $i=0$ and make all keys eligible (no anchor pile). 
Whenever a nonempty pile $P_i$ of size at most $s$
is chosen as the anchor pile,
use Lemma~\ref{lem:skip} as the fixed eligibility predicate. Test relative ranks $j=1,2,\ldots$ in order. For each $j$, enumerate strict decreases of the top on original input prefixes, using the recursive query and an empty artificial prefix. Starting at prefix zero with top $+\infty$, test whether the final top is strictly smaller than the current top. If so, binary search for the first longer prefix with a strictly smaller top. Monotonicity ensures correctness, and ordinary patience sorting identifies this position as exactly the next member of the relative pile $j$. Update the current top and repeat, abandoning the test after $s+1$ members. Each discovered member costs ${\rm O}(\log(2n))$ queries; an unsuccessful final test costs one.

\begin{figure}[tbp]
\centering
\begin{tikzpicture}[every node/.style={font=\small},>=stealth]
\draw[->] (-.6,0)--(11,0) node[right]{rank};
\foreach \x/\lab in {0/i,2/{i+1},6/{i+j},9.8/{i+d}} {\draw (\x,.12)--(\x,-.12) node[below]{$\lab$};}
\draw[very thick] (0,.08)--(0,.7); \node[above] at (0,.7){anchor};
\draw[very thick] (6,.08)--(6,.7); \node[above] at (6,.7){next small pile};
\draw[->,thick] (.4,.35)--(5.6,.35) node[midway,above]{test in order};
\draw[<->] (2,-.65)--(9.8,-.65) node[midway,below]{at most $d$ candidate ranks};
\node at (3.9,-1.55){Intermediate piles have more than $s$ members.};
\end{tikzpicture}
\caption{A successful jump from anchor rank $i$ to a nonempty pile at rank $i+j$, with $j\le d$. If the selected pile is empty, the algorithm terminates instead.}
\label{fig:skip}
\end{figure}
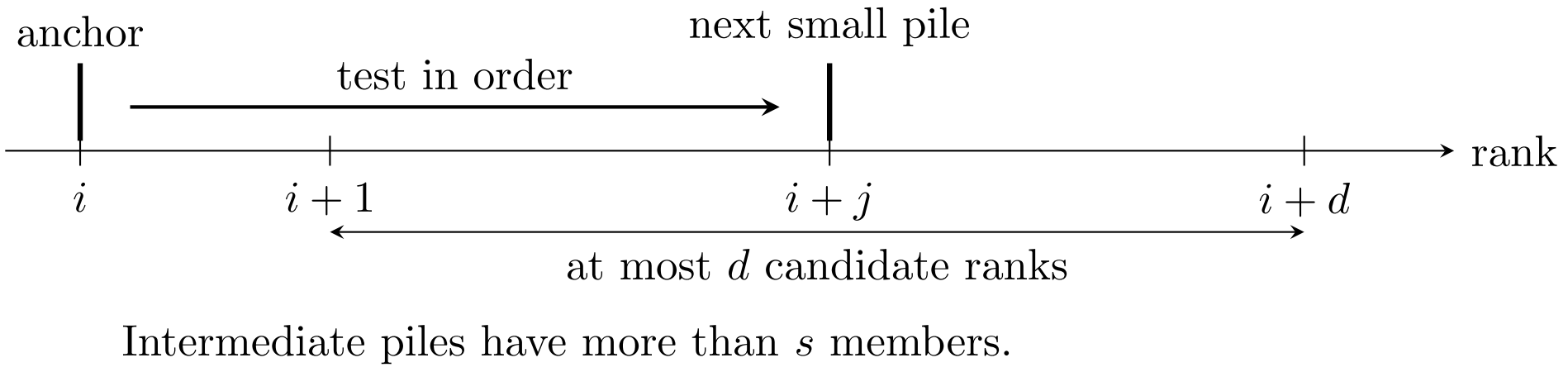

The first pile of size at most $s$ is either empty, in which case return $i+j-1$, or nonempty, in which case replace the anchor pile by it and advance $i$ by $j$. At most $\lfloor n/(s+1)\rfloor$ piles can have more than $s$ members, so the largest requested relative rank is at most
\[
d=\lfloor n/(s+1)\rfloor+1.
\]
Figure~\ref{fig:skip} illustrates the rank traversal.
The anchor pile is sorted by position, so eligibility is testable by binary search for its last preceding member in ${\rm O}(\log(2n))$ time and constant additional words. Each absolute rank is tested at most once. The total number of discovered members is at most $\sum_i\min(|P_i|,s+1)\le n$, and there are at most $n+1$ rank tests. Thus there are ${\rm O}(n\log(2n))$ top queries in total, outside the recursive tree.

\section{Complexity}
\begin{theorem}
\label{thm:length}For every fixed $c>0$, LIS length is computable deterministically in $n^{2+4c^2+{\rm o}(1)}$ time using
\[
{\rm O}\left(\frac{\sqrt n(\log n)^{3/2}}{2^{c\sqrt{\log n}}}\right)={\rm o}(\sqrt n)
\]
bits of workspace.
\end{theorem}
\begin{proof}
Put $f=2^{c\sqrt{\log n}}$ and $s=\max(1,\lfloor\sqrt n/f\rfloor)$. For sufficiently large $n$, $s=\Theta(\sqrt n/f)$. \paragraph{Branching factor.} The maximum branching factor is at most $2\lceil(d+1)/s\rceil={\rm O}(f^2)$, whose logarithm is at most $2c\sqrt{\log n}+{\rm O}(1)$. \paragraph{Depth.} The depth is at most $\lceil\log(n/s^2)\rceil=2c\sqrt{\log n}+{\rm O}(1)$. \paragraph{Number of nodes.} Hence a query has at most
\[
\left(2^{2c\sqrt{\log n}+{\rm O}(1)}\right)^{2c\sqrt{\log n}+{\rm O}(1)}\!\cdot {\rm O}(\sqrt{\log n}+1) = n^{4c^2+{\rm o}(1)}
\]
nodes, since $2^{{\rm O}(\sqrt{\log n})}=n^{{\rm o}(1)}$. \paragraph{Time per node and overall time.} Internal local work is ${\rm O}(s+d)$ per node (linear in $s$ per block); leaf work is ${\rm O}(n\log(2n))$ by the truncated scans described above. Each query thus takes $n^{1+4c^2+{\rm o}(1)}$ time. Multiplying by the ${\rm O}(n\log(2n))$ outer queries gives the stated time bound in the model of Section~2.

\paragraph{Space.} Each frame and leaf uses ${\rm O}(s\log n)$ bits, as do the old and prospective outer piles. Total space is ${\rm O}(s\log n(1+\log f))$. Bounded exceptional input sizes can be handled directly.
\end{proof}
\paragraph{Corollary.} For every fixed $\varepsilon>0$, LIS length is computable in ${\rm O}(n^{2+\varepsilon})$ time and ${\rm O}(\sqrt n/2^{\delta\sqrt{\log n}})={\rm o}(\sqrt n)$ bits of workspace for some constant $\delta>0$. Choose $c>0$ with $4c^2<\varepsilon$ to absorb the ${\rm o}(1)$ in the time exponent, and any $0<\delta<c$ to absorb the polylogarithmic space factor.

\section{Outputting a longest increasing subsequence}\label{sec:witness}
We present two reconstruction methods. The first is a simple endpoint-based method that illustrates how to recover a witness but costs an extra factor of $n$. The second, balanced reconstruction, achieves the main theorem's time bound and is our final algorithm.
\subsection{Simple endpoint-based reconstruction}
The length algorithm also identifies an endpoint of an LIS without increasing its asymptotic resource bounds. Indeed, if the final empty pile has absolute rank $\ell+1$, the preceding nonempty pile of rank $\ell$ has already been tested (or is the current anchor pile). In the former case, its final top can be obtained by one additional top query while testing that rank, even if enumeration was abandoned after $s+1$ members. Retain this top whenever a nonempty pile is tested. In the latter case, its final top is the last member of the anchor pile. For $\ell=0$, there is no endpoint to return. Thus the algorithm returns both $\ell$ and an original position $q$ of rank $\ell$, using at most a constant number of additional words and at most one additional query per rank test. The time bound of Theorem~\ref{thm:length} is unchanged.

\begin{theorem}
For every fixed constant $c>0$, an LIS can be output in input order in deterministic $n^{3+4c^2+{\rm o}(1)}$ time using
\[
{\rm O}\left(\frac{\sqrt n(\log n)^{3/2}}{2^{c\sqrt{\log n}}}\right)
\]
bits of working space, with a read-only input and a write-only output.
\end{theorem}
\begin{proof}
First consider output in reverse order. If the current sequence has LIS length $\ell>0$, use the augmented length algorithm to find an endpoint $q$ of rank $\ell$ and output that key (or its original position). Restrict the current sequence to keys preceding $q$ and strictly smaller than its key. The restricted sequence has LIS length exactly $\ell-1$: deleting the endpoint from a longest chain ending at $q$ proves the lower bound, and appending $q$ to any increasing chain of the restricted sequence proves the upper bound. Repeat until the sequence is empty. By induction, the emitted keys in reverse order form an LIS.

No restricted sequence is materialized. Each iteration maintains a prefix boundary and a key upper bound; successive restrictions can be represented by replacing these two parameters. Run the length algorithm on the original index domain of size $n$, retaining only keys satisfying this fixed restriction in addition to its usual eligibility predicate. Testing the restriction costs constant time per original key; the disjoint-pile counting arguments use at most $n$ retained keys, and the recursive depth and leaf scanning bounds remain valid with the original parameter $n$. Each call therefore obeys the original time and workspace bounds. There are at most $n$ nonempty iterations and one final termination test. Working memory is reused between calls, with only ${\rm O}(\log n)$ additional bits; no output is read back or stored.

To output in input order, perform the same procedure on the sequence in reversed position order with its total key order reversed as well. Increasing subsequences in this transformed sequence correspond bijectively to reversed increasing subsequences of the original sequence. Reverse-order emission in the transformed sequence is consequently forward-order emission in the original sequence. Both transformations are implemented by index arithmetic and reversing comparisons, including the tie-breaking order for repeated keys. This requires no additional input storage.
\end{proof}
\paragraph{Corollary.} For every fixed $\varepsilon>0$, an LIS can be output in ${\rm O}(n^{3+\varepsilon})$ time and ${\rm O}(\sqrt n/2^{\delta\sqrt{\log n}})={\rm o}(\sqrt n)$ bits of workspace for some constant $\delta>0$. This follows by choosing $4c^2<\varepsilon$ and $0<\delta<c$ as in the length-only corollary. The next subsection removes this additional factor.

\subsection{Balanced reconstruction by streaming pile intersections}
The additional factor of $n$ in the preceding theorem can be avoided. We use the forward/backward divide-and-conquer framework of~\cite{lisjournal}, but replace its search for a small central pile by a streaming intersection of two arbitrary piles.

\begin{lemma}\label{lem:enum}
On an original interval of length $m$ with a fixed key-range restriction, any specified pile can be enumerated in either position order in $m^{2+4c^2+{\rm o}(1)}$ time. Its suspended enumerator occupies ${\rm O}(s\log n)$ bits, where $s=\max(1,\lfloor\sqrt m/2^{c\sqrt{\log m}}\rfloor)$ and $n$ is the original input length. A running top query uses ${\rm O}(s\log n(1+\log(m/s^2)))$ bits.
\end{lemma}
\begin{proof}
Use the outer traversal of Section~\ref{sec:outer}, with $d=\lfloor m/(s+1)\rfloor+1$, until the desired rank $k$ satisfies $1\le k-i\le d$, where $i$ is the anchor rank (initially zero). Whenever $k-i>d$, a small pile is found among the next $d$ ranks, strictly before $k$. If an empty pile intervenes, the desired pile is empty. By Lemma~\ref{lem:skip}, the desired pile is now the relative pile $k-i$ of the eligible sequence. All preparation tests concern distinct absolute ranks below $k$ and discover at most $m$ members in total.

Enumerate the desired pile without the cutoff at $s+1$ members. Forward enumeration is exactly the strict-decrease search of Section~\ref{sec:outer}. For reverse enumeration, let $F(t)$ be the relative top on the eligible prefix ending at raw position $t$ (using local interval coordinates). Start with $r=m$. Compute $v=F(r)$; if it is infinite, stop. Otherwise binary search on $[1,r]$ for the first $t$ with $F(t)\le v$. Monotonicity and $F(r)=v$ imply equality at that position. Distinct keys guarantee that $t$ is the last update position at or before $r$. Return $t$ and set $r=t-1$. Repeat. Each member requires ${\rm O}(\log(2m))$ queries in either direction; the entire preparation and enumeration use ${\rm O}(m\log(2m))$ queries by disjointness of piles. Theorem~\ref{thm:length}'s query analysis with parameter $m$ gives the time bound. The fixed key-range test takes constant time per scanned key; scans and recursive splits use the raw interval, not the entire input or a materialized filtered sequence. A suspended enumerator retains only its small anchor pile, rank, cursor and current key; it suspends only between completed queries. Original indices require ${\rm O}(\log n)$ bits per word. The query stack bound follows as before. Bounded small intervals are handled directly.
\end{proof}

\begin{theorem}\label{thm:fastoutput}
For every fixed $c>0$, an LIS can be output in input order in deterministic $n^{2+4c^2+{\rm o}(1)}$ time using
\[
{\rm O}\left(\frac{\sqrt n(\log n)^{3/2}}{2^{c\sqrt{\log n}}}\right)
\]
bits of working space, with read-only input and write-only output.
\end{theorem}
\begin{proof}
\textit{Existence of a middle element.} Consider a restricted interval whose LIS length $\ell$ is known. Return immediately for $\ell=0$. Otherwise let $k=\lceil\ell/2\rceil$. Write $L(q)$ and $R(q)$ for the maximum lengths of increasing chains ending and starting at position $q$, respectively, in this restricted sequence. The forward pile $P_k$ consists of positions with $L(q)=k$; the backward pile $Q_{\ell-k+1}$ consists of positions with $R(q)=\ell-k+1$. The latter is an ordinary pile in the sequence with both position and total key order reversed. The intersection is nonempty: the $k$th member of any LIS belongs to both piles, since a longer prefix or suffix chain would contradict maximality. Conversely, concatenating maximum chains through any common member gives an LIS.

\textit{Finding the intersection.} Prepare the two enumerators of Lemma~\ref{lem:enum}. Enumerate the forward pile in reverse position order and the backward pile in forward order in the transformed sequence, so both streams are decreasing in original position. Compare their current positions, advancing the stream with the larger position until equality. A common position $q$ is guaranteed. The two suspended states occupy ${\rm O}(s\log n)$ bits together; their top queries run sequentially and reuse one scratch area. Hence the middle-element search has the time and space bounds of the lemma.

\textit{Recursive output and correctness.} Release both enumerators and their scratch space. Recursively output the LIS on the portion strictly before $q$ with keys below its key, output $q$, then recursively output the LIS on the portion strictly after $q$ with keys above its key. Inherit the previous key bounds in both children. The two lengths are exactly $k-1$ and $\ell-k$, by the definitions of $L(q)$ and $R(q)$. Induction proves correctness and forward output order. The initial length is computed once by Theorem~\ref{thm:length}; no further length computations are necessary.

\textit{Time analysis.} Every nonempty child has LIS length at most $\lfloor\ell/2\rfloor$, so reconstruction has depth ${\rm O}(\log(2n))$. At each depth the raw intervals are disjoint, with lengths $m_j$ summing to at most $n$. For any fixed $\eta>0$, putting $p=2+4c^2+\eta$, the time at that depth is ${\rm O}(\sum_j m_j^p)={\rm O}(n^p)$. The logarithmic number of levels is absorbed into $n^{{\rm o}(1)}$ (choose $\eta$ arbitrarily small). As in Section~2, key comparisons and word operations take constant time, including
for local intervals of the original input.

\textit{Space analysis.} Each reconstruction frame stores only interval endpoints, key bounds, the known length, the pivot and a constant amount of control information, for a total of ${\rm O}(\log^2 n)$ bits. At most one middle-element search is active at a time. Its workspace is ${\rm O}(s(m)\log n(1+\log(m/s(m)^2)))$. For sufficiently large $m$, this is bounded by ${\rm O}(\sqrt m\,2^{-c\sqrt{\log m}}\log n\,(1+\sqrt{\log m}))$. The function $\sqrt m\,2^{-c\sqrt{\log m}}$ is eventually increasing (write its base-two logarithm as $x/2-c\sqrt{x}$, where $x=\log m$; its derivative is $1/2-c/(2\sqrt{x})>0$ for $x>c^2$); bounded exceptional sizes use ${\rm O}(\log n)$ bits. Thus the maximum over $m\le n$, plus the reconstruction stack, satisfies the asserted space bound.
\end{proof}
\paragraph{Corollary.} For every fixed $\varepsilon>0$, both LIS length and an LIS witness are computable in ${\rm O}(n^{2+\varepsilon})$ time and ${\rm O}(\sqrt n/2^{\delta\sqrt{\log n}})={\rm o}(\sqrt n)$ bits of workspace for some fixed $\delta>0$, by choosing $4c^2<\varepsilon$ and $0<\delta<c$.

\section{Consequences and limitations}
The standard reduction from LCS to LIS
\cite{hunt1977}
transforms an LCS
instance on two length-$N$ strings into a strict LIS
instance with at most $N^2$ keys;
enumerate matches by increasing first index and decreasing second index within each group. Access can be simulated by rescanning the strings. Substitution yields space ${\rm O}(N(\log N)^{3/2}/2^{c\sqrt{2\log N}})$ for every fixed $c>0$, and $N^{4+8c^2+{\rm o}(1)}$ time if matching pairs are supplied by a constant-time access oracle. For explicitly given strings, rescanning to simulate each access incurs at most an additional ${\rm O}(N^2)$ factor, preserving polynomial time. 
However, the same space bound can be achieved more
efficiently by the direct LCS algorithm of~\cite{lcs}.
Indeed, for any fixed $\alpha>0$, choosing the block size
$B=\lceil N/2^{\alpha\sqrt{\log N}}\rceil$ in that algorithm
gives a recursion tree with branching factor ${\rm O}(N/B)$
and depth $\alpha\sqrt{\log N}+{\rm O}(1)$.
The number of nodes is therefore
$N^{\alpha^2+{\rm o}(1)}$.
Since each leaf takes ${\rm O}(BN)$ time and each recursion
frame uses ${\rm O}(B\log N)$ bits, the resulting algorithm
runs in $N^{2+\alpha^2+{\rm o}(1)}$ time using
\[
{\rm O}\left(
\frac{N(\log N)^{3/2}}{2^{\alpha\sqrt{\log N}}}
\right)
\]
bits of working space.
Setting $\alpha=\sqrt{2}c$ gives the same asymptotic
space bound as the LIS-based approach, but with a
smaller time exponent.
Thus, our result does not improve the existing
time--space bounds for LCS.

Although our algorithms break the square-root space
barrier, they do not achieve a polynomial-factor
improvement over $\sqrt n$ in the space bound.
A natural next question is whether LIS can be solved
in polynomial time using ${\rm O}(n^{1/2-\varepsilon})$ bits
of working space for some fixed $\varepsilon>0$.
As pointed out in~\cite[Section 7]{lisjournal},
such an improvement would also have implications
for LCS. Indeed, by the above reduction, it would
yield a polynomial-time algorithm for LCS using
${\rm O}(N^{1-2\varepsilon})$ bits of working space.
Thus, obtaining a polynomial-factor improvement
over the square-root space bound for LIS would
also break the linear-space barrier for LCS by
a polynomial factor.
\

\end{document}